\documentclass[%
 reprint,
 amsmath,amssymb,
 aps,nofootinbib,
  onecolumn,
]{revtex4-1}
\usepackage[usenames, dvipsnames]{color}
\usepackage[autostyle]{csquotes}
\usepackage{pgfplots}
\usepackage{tikz}
\usepackage{graphicx}
\usepackage{dcolumn}
\usepackage{bm}
\usepackage{enumerate}

\begin{document}

\preprint{APS/123-QED}

\title{Entropy Production in Affine Inflation}

\author{Hemza Azri $^{\it \bf a}$}
 \email{hmazri@uaeu.ac.ae}
\author{Salah Nasri $^{\it \bf a, \bf b}$}%
 \email{snasri@uaeu.ac.ae}
\affiliation{$^{\bf \it a}$%
 Department of Physics, United Arab Emirates University, UAE
}%
\affiliation{$^ {\bf \it b}$ International Center for Theoretical Physics, Trieste, Italy}




\date{\today}







\date{\today}







\date{\today}

\begin{abstract}
 Multiple scalar fields nonminimally interacting through pure affine gravity are considered to generate primordial perturbations during an inflationary phase. The couplings considered give rise to two distinct sources of entropy perturbations that may not be suppressed in the long wavelength limit. The first is merely induced by the presence of more than one scalar and arises even in the minimal coupling limit. The second source however is restricted to nonminimal interaction. Unlike the case of metric gravity, and due to the absence of anisotropic stresses, the second source disappears for single scalar, showing that nonminimal couplings become relevant to non-adiabatic perturbations only when more than one scalar field are considered. Hence the notion of adiabaticity is not affected by the transition to minimal coupling contrary to the metric gravity case where it is confused by changing the frames. Precise data that might be able to neatly track different sources of isocurvature modes, if any, must not only distinguish between different models of inflation but also determine the most viable approach to gravity which underlies the inflationary dynamics itself.
 
\begin{description}
\item[Keywords]
Inflation; Entropy perturbation; Non-adiabatic pressure; Isocurvature; Affine gravity. 
\end{description}
\end{abstract}

\maketitle


\section{\label{sec:intro} Introduction}
\label{sec: introduction}
One of the primary interest of the very early universe theories lies in understanding the origin of structure. Inflationary cosmology serves  as a relevant mechanism in which the vacuum fluctuations, in an early phase of very rapid accelerated expansion, swept up to large scales and acts as seeds of structure formation later on. In its simplest realization by a slowly rolling single scalar field, inflation provides us with a nearly scale invariant spectrum of Gaussian, adiabatic density perturbations that fit observational constraints \cite{planck2018}. However, since there are many models of inflation where predictions are relatively in agreement with observations, a specific model then is still yet to be determined with the help of a future more precise data \cite{martin}. 

Various possible theoretical realizations, such as incorporating more than one scalar fields to drive inflation have opened the question about the adiabatic character of the early cosmological perturbations. It turned out that multiple fields can generically lead to isocurvature (non-adiabatic) perturbations \cite{wands0, wands1, wands2}. While in single field models non-adiabatic (entropy) perturbation modes are merely suppressed in the long wavelength limits (super-horizon scales), in multiple fields models however, these modes can in principle amplify the curvature perturbations and alter its evolution even after they have crossed outside the horizon (Hubble radius). 

Multiple fields can enter the gravitational action in various ways, the simplest is to minimally interact where all the fields enjoy a canonical kinetic term, and another way is by interacting directly with  the spacetime curvature (nonminimal coupling). In both cases\footnote{There is still  an other possibility in which the fields gain a non-canonical kinetic terms. This case would lead to models such as $k$-Inflation which is beyond the scope of  the present paper \cite{k-inflation}.} generating entropy perturbations that must not be suppressed on super horizon scales is inevitable \cite{kaiser0, kaiser1, sasaki}. In fact,  not only multiple fields but even one single field can source non-adiabatic perturbation if it is nonminimally coupled to gravity \cite{kaiser0, kaiser1}.  This surely indicates that nonminimal couplings, at least in metric gravity, are relevant to isocurvature and thus play a role in affecting the evolution of the curvature perturbations (nevertheless, an attempt to exclude single-filed models  with such a feature  have been considered  in \cite{sasaki} by using  different geometric formulation.) However, we notice that the last conclusion may lead to some confusion concerning the notion of adiabaticity when swiching back to Einstein frame in which the nonminimal coupling interaction disappears. In fact, at least for single field case, while the gauge invariant curvature perturbation is conserved in Einstein frame, in Jordan frame it becomes time-dependent although the two frames are physically equivalent. This issue is caused by the \enquote{metric} conformal transformation which is a feature of \textit{metric gravity}.
To that end, it maybe more viable to rather consider a metric-less gravity for inflation itself. Purely affine theory of gravity, based solely on affine connection with no notion of metric, supports already scalar fields with non-vanishing potentials and stands viable for inflationary dynamics \cite{affine-inflation, induced-affine-inflation}. Dynamics of nonminimally and minimally coupled scalar fields are described by two pure affine invariant actions that could be transformed to each other using only simple field redefinition. Since the geometric part is not altered by this redefinition, important quantities such as the Hubble parameter, and then the curvature perturbations are not subjected to changing \cite{azri-review}.

In the present paper we thoroughly study non-adiabatic perturbations during inflation in the context of affine gravity. The main goal is to track the possible sources of entropy perturbations that may not be suppressed in long wavelength limit. Our framework will be based on a primary affine action in which multiple scalars are nonminimally coupled to gravity via the affine curvature. The linearity of the curvature with respect to the affine connection results in a generalized energy-momentum free of any additional terms that represent anisotropic pressure, such that at first order in perturbation only isotropic components including momentum flow contribute to the dynamics of the perturbations. This compact form leads to significant simplifications compared to metric gravity, where the Bardeen potentials in Newtonian gauge are equal even in the case of nonminimal coupling. In this case, non-adiabatic pressure of the system which may not be suppressed on super-horizon scales will appear in terms of two distinct quantities where one is sourced by the presence of more than one scalar which is a generic source that holds even in metric gravity, but the second is related to nonminimal coupling. It turns out that the second source vanishes in the single field-case leaving us with the conclusion that nonminimal couplings do not contribute in entropy production unless multiple fields are present. It is only from future more precise data that one will be able  to discriminate between theories with or without entropy perturbations by analyzing the power spectra of the cosmic microwave background anisotropies and polarization, and tracking any hints of isocurvature modes. The present paper is considered as a generic framework and model-independent formulation of entropy perturbations in (affine) inflation, and a future work will be devoted to an application, with some specific models, that runs along the present results.     

The paper is organized as follows, the next section will be devoted to an overview of pure affine gravity with multiple scalars. Since it may not be familiar to the reader, we will bring detailed calculation for the derivation of the gravitational equations and the evolution of the scalar fields. In Sec \ref{sec: field fluctuations and entropy perturbation} we tackle the scalar perturbations and study their evolution and see how they look like compared to the case of metric gravity. Finally we derive the non-adiabatic pressures sources responsible for entropy perturbations. In Sec \ref{sec: conclusion} we summarize the main results and conclude.

\section{Multiple fields in affine gravity: an overview}
\subsection{Nonminimal coupling and field equations}
In purely metric theories of gravity (general relativity and its modifications) the interaction of matter fields with gravity is trivially performed by generalising related field theory Lagrangian densities in flat space such that the flat Minkowski metric is replaced by a curved spacetime metric tensor. The later is essential in contracting any matter or geometric tensor fields that allows finally for the construction of covariant actions. In the absence of any source of matter fields as well as vacuum energy, the gravitational field equations in free space is easily derived from Einstein-Hilbert action. However, it has been known that there is no fundamental principle that stands against extending spacetime structure itself by an ingredient more fundamental than the metric tensor. In fact, the major achievement of relativistic gravity is to introduce the concept of the connection that enables defining infinitesimal displacement of tensor fields in the curved background. Gravitational strengths then, would be measured by the curvature of this connection which in turn has no \textit{a priori} relation with the metric.  To that end, one would alternatively consider two possible approaches: (\textit{i}) \textit{metric-affine} \cite{hehl, karahan} where both metric and affine connection are introduced independently,  (\textit{ii}) \textit{purely affine} \cite{affine-inflation, induced-affine-inflation, azri-review, kijowski, kijowski1, poplawski, azri-thesis, azri-separate, azri-eddington, azri-induced} in which no metric tensor is considered \textit{a priori} but only an affine connection as a fundamental field. In this paper we are considering the second approach particularly the approach with nonminimal coupling provided in \cite{affine-inflation, induced-affine-inflation, azri-review}.     

In the absence of metric tensor, the number of quantities that one could consider are less than that of the metric case, recalling  that scalars formed by contractions (using metric) are not allowed in the first place. One could however consider scalar fields $\phi^{1}, \dots, \phi^{N}$ and their derivatives as well as associated potential $V(\phi^{1}, \dots, \phi^{N})$. In the geometric sector we mainly have a symmetric connection $\Gamma_{\mu\nu}^{\lambda}$ and the associated curvature or Ricci tensor $R_{\mu\nu}(\Gamma)$ which for simplicity will be taken symmetric too (only symmetric part is taken.)  Despite its simple structure, forming a familiar polynomial gravitational action is not trivial; this clearly stems from the absence of some essential covariant ingredients such as field kinetic terms which requires a metric tensor  (see \cite{oscar1, oscar2} for attempts to construct  polynomial affine actions). 

Nevertheless, pure affine actions can still be constructed not as polynomials but in terms of volume measures. In fact, 
one can form the following diffeomorphism invariant action \cite{azri-review}    
\begin{eqnarray}
\label{action1}
S[\Gamma, \phi^{1}, \dots, \phi^{N}]=
\int d^{4}x \frac{\sqrt{ |f(\phi^{1}, \dots, \phi^{N} )R_{\mu\nu}(\Gamma)- \delta_{ab}\nabla_{\mu}\phi^{a} \nabla_{\nu}\phi^{b} |}}
{V(\phi^{1}, \dots, \phi^{N})},
\end{eqnarray}
where the matter  fields indices run as $a,b=1, \dots, N$.

The  spacetime coordinate dependent function $f(\phi^{1}, \dots, \phi^{N} )$ represents  the nonminimal coupling function to the curvature  and it can be considered as a varying mass, which reduces to the Planck mass in the case of minimal coupling to gravity. For a single field interacting nonminimally, this function would generically take the form $f(\phi)=M^{2}_{Pl}+ \xi \phi^{2}$ (see \cite{affine-inflation} concerning the affine approach of this case.) An interesting feature of the last action that metric gravity theories do not enjoy is the appearance of the potential energy in a denominator rather than in a separate term. If the total potential vanishes the action suffers from singularity, a feature that shows how potentials are crucial in these type of  models and this is what inflation requires already at the first place. This would imply that when the scalar field's potential enjoys symmetry-breaking solutions with some nonzero vacuum expectation values, one should certainly improve the potential with a nonzero constant term which will describe a possible cosmological constant that prevents the action from  becoming  singular in the vacuum \cite{induced-affine-inflation}.   

The gravitational equations are derived by varying action (\ref{action1}) with respect to the affine connection. The later appears only in the curvature in terms of its first covariant derivative and quadratic forms that renders variation of curvature more compact as
\begin{eqnarray}
\delta R_{\mu\nu}(\Gamma)=
\nabla_{\lambda}\left( \delta \Gamma_{\mu\nu}^{\lambda} \right)
- \nabla_{\nu}\left( \delta \Gamma_{\lambda\mu}^{\lambda} \right).
\end{eqnarray}
This would easily lead to the following infinitesimal variation of our action    
\begin{eqnarray}
\delta_{\Gamma} S=
\frac{1}{2}\int d^{4}x
\Bigg\{ \nabla_{\mu} \left(f \frac{\sqrt{|K(\Gamma, \phi)|}}{V(\phi)}(K^{-1})^{\alpha\mu} \right)\delta_{\lambda}^{\beta}
- \nabla_{\lambda} \left(f \frac{\sqrt{|K(\Gamma, \phi)|}}{V(\phi)}(K^{-1})^{\alpha\beta} \right)
\Bigg\}\delta \Gamma_{\alpha\beta}^{\lambda}, \nonumber
\\
\end{eqnarray}
where we have used for brevity the following \enquote{Kinetic} part of the gravitational and scalar  field sectors that enters the action as a tensor
\begin{eqnarray}
K_{\mu\nu}(\Gamma, \phi^{1}, \dots, \phi^{N})=
f(\phi^{1}, \dots, \phi^{N} )R_{\mu\nu}(\Gamma)- \delta_{ab}\nabla_{\mu}\phi^{a} \nabla_{\nu}\phi^{b}.
\end{eqnarray}
The action stays stationary under variation when the following dynamical equations are satisfied
\begin{eqnarray}
\nabla_{\mu} \left(f \frac{\sqrt{|K(\Gamma, \phi)|}}{V(\phi)}(K^{-1})^{\alpha\mu} \right)\delta_{\lambda}^{\beta}
- \nabla_{\lambda} \left(f \frac{\sqrt{|K(\Gamma, \phi)|}}{V(\phi)}(K^{-1})^{\alpha\beta} \right)=0,
\end{eqnarray}
which takes the  simple final form
\begin{eqnarray}
\label{dynamical equation}
\nabla_{\lambda} \left(f(\phi^{1}, \dots, \phi^{N} ) \frac{\sqrt{|K(\Gamma, \phi)|}}{V(\phi^{1}, \dots, \phi^{N} )}(K^{-1})^{\alpha\beta} \right)=0.
\end{eqnarray}
This is the main equation that will basically govern the affine dynamics. As we shall see later, two important consequences will arise form this equation, the first and most crucial is \textit{generating} a metric tensor, whereas the second is the gravitational field equations in terms of this metric. For the moment, it is important to notice that like the primary action (\ref{action1}) the last dynamical equation does not involve any and refer to any metric tensor.  

Before writing the gravitational equations, let us first focus on the dynamics of the scalar fields. These are described and given by their equations of motion derived by varying the action with respect to the scalar fields themselves where in this case
\begin{eqnarray}
\delta_{\phi} S=
\int d^{4}x \Bigg \{
\frac{1}{2}\frac{\partial f}{\partial \phi^{a}} \frac{\sqrt{|K(\Gamma, \phi)|}}{V(\phi)}(K^{-1})^{\alpha\beta}R_{\alpha\beta}(\Gamma)
+\partial_{\alpha}\left( \frac{\sqrt{|K(\Gamma, \phi)|}}{V(\phi)}(K^{-1})^{\alpha\beta}\partial_{\beta}\phi^{a}  \right)  
 -\frac{\sqrt{|K(\Gamma, \phi)|}}{V^{2}(\phi)} \frac{\partial V}{\partial \phi^{a}} \Bigg\} \delta \phi^{a}
\end{eqnarray}
which leads to the equations of motion
\begin{eqnarray}
\label{scalar-eom1}
\partial_{\alpha}\left( \frac{\sqrt{|K(\Gamma, \phi)|}}{V(\phi)}(K^{-1})^{\alpha\beta}\partial_{\beta}\phi^{a}  \right)
+\frac{1}{2}\frac{\partial f}{\partial \phi^{a}} \frac{\sqrt{|K(\Gamma, \phi)|}}{V(\phi)}(K^{-1})^{\alpha\beta}R_{\alpha\beta}(\Gamma)
-\frac{\sqrt{|K(\Gamma, \phi)|}}{V^{2}(\phi)} \frac{\partial V}{\partial \phi^{a}} 
=0.
\end{eqnarray}
Though appears complicated, this is nothing but the equation of motion that govern the dynamics of the scalar field $\phi^{a}$ in a curved affine background. Again, before generating the metric and writing this equation in a \enquote{familiar} form, one should notice its independence of the metric.

Let us now return to the gravitational sector and examine equation (\ref{dynamical equation}). The first result we can obtain from this equation is that the affine connection that has been taken arbitrary (but symmetric) in the action obeys now a constraint that reduces it to the Levi-Civita connection of an invertible tensor that we shall denote as $g_{\mu\nu}$.  In other words, this tensor is generated as a solution to equation (\ref{dynamical equation}) by setting
\begin{eqnarray}
\label{metric}
f(\phi^{1}, \dots, \phi^{N} ) \frac{\sqrt{|K(\Gamma, \phi)|}}{V(\phi^{1}, \dots, \phi^{N} )}(K^{-1})^{\alpha\beta}=
M^{2}_{\text{Pl}}\sqrt{|g|}(g^{-1})^{\alpha\beta},
\end{eqnarray}
where $M_{\text{Pl}}$ is simply the Planck mass that balances the dimensionality in the last equation since the nonminimal coupling function $f(\phi^{1}, \dots, \phi^{N} )$ has a dimension of mass squared. It could have been taken as an arbitrary mass, however consistency with Einstein field equations in vacuum implies that this mass must coincide with the Planck mass \cite{affine-inflation}.

Now with the aid of the last equation, the dynamical equation (\ref{dynamical equation}) becomes a gravitational field equations with a metric tensor $g_{\mu\nu}$ compatible with the connection\footnote{In general, These relations lead to $\nabla_{\lambda} (\sqrt{|g|}(g^{-1})^{\alpha\beta})=0$, but since the affine connection is taken symmetric one easily obtains the compatibility condition (\ref{compatibility}). For the metric to be physical, only those configurations $(\Gamma,\phi^{i})$ for which $K_{\mu\nu}$ has the signature $(-,+,+,+)$ are considered \cite{kijowski}.}, thus
\begin{eqnarray}
\label{equations in terms of ricci}
&&K_{\mu\nu}(g, \phi^{1}, \dots, \phi^{N})=\frac{M^{2}_{\text{Pl}} V(\phi^{1}, \dots, \phi^{N} )}{f(\phi^{1}, \dots, \phi^{N} )}  g_{\mu\nu}, \\ 
&&\nabla_{\lambda} g_{\mu\nu}=0.
\label{compatibility}
\end{eqnarray}
This couple of equations shows the interesting transition from the pure affine dynamics of the system to the metrical structure where the later arises only \textit{a posteriori} and not imposed from scratch. With this metric, lowering and raising indices as well as contractions become possible, and finally one can form the Einstein tensor, and equations (\ref{equations in terms of ricci}) takes the form
\begin{eqnarray}
\label{eom-gravity}
f(\phi^{1}, \dots, \phi^{N} )G_{\mu\nu}=
\delta_{ab}\nabla_{\mu}\phi^{a} \nabla_{\nu}\phi^{b}
-\frac{1}{2}\delta_{ab}\nabla^{\lambda}\phi^{a} \nabla_{\lambda}\phi^{b} g_{\mu\nu}
-\frac{M^{2}_{\text{Pl}} V(\phi^{1}, \dots, \phi^{N} )}{f(\phi^{1}, \dots, \phi^{N} )} g_{\mu\nu}.
\end{eqnarray}
It is clear that for the minimal coupling limit where $f = M^{2}_{\text{Pl}}$, the last field equations get reduced to Einstein field equations with scalar fields in a familiar form. Thus, the pure affine action gives rise to Einstein equations where spacetime curvature is sourced by the generalised energy-momentum tensor of the form
\begin{eqnarray}
\label{energy-momentum tensor}
\mathcal{T}_{\mu\nu}=
\frac{1}{f(\phi^{1}, \dots, \phi^{N} )}  \left\{
\delta_{ab}\nabla_{\mu}\phi^{a} \nabla_{\nu}\phi^{b}
-\frac{1}{2}\delta_{ab}\nabla^{\lambda}\phi^{a} \nabla_{\lambda}\phi^{b} g_{\mu\nu}
-\frac{M^{2}_{\text{Pl}} V(\phi^{1}, \dots, \phi^{N} )}{f(\phi^{1}, \dots, \phi^{N} )} g_{\mu\nu}
\right\}.
\end{eqnarray}
One might easily notice the difference between this tensor and the energy-momentum tensor that arises from nonminimal coupling of pure metric gravity. The crucial difference relies on the absence of the terms proportional to $\nabla_{\mu}\nabla_{\nu}f - g_{\mu\nu} \Box f$ which appear in metrical gravity due to the nonlinearity of Einstein-Hilbert action. These terms are the sources of the so called \textit{anisotropic pressure} and their presence certainly affects the scalar fields dynamics. Among its effects is the contribution to generating entropy perturbations through non-adiabatic pressure even for a single scalar field \cite{kaiser0, kaiser1}, this  however is prevented  as we shall see in Sec \ref{sec: field fluctuations and entropy perturbation} when studying scalar perturbations.

The same for the evolution of the scalar fields, using the generated metric (\ref{metric}), the scalar field equations of motion (\ref{scalar-eom1}) take the form
\begin{eqnarray}
\label{eom-field dynamics}
\Box \phi^{a} -\frac{\partial V}{\partial \phi^{a}} +\frac{1}{2} \frac{\partial f}{\partial \phi^{a}}R(g) +\psi(\phi^{1}, \dots, \phi^{N})=0,
\end{eqnarray}
where
\begin{eqnarray}
\label{psi}
\psi(\phi^{1}, \dots, \phi^{N})=
\left(1-\frac{M^{2}_{\text{Pl}}}{f}\right) \frac{\partial V}{\partial \phi^{a}}
-\frac{1}{f}\frac{\partial f}{\partial \phi^{a}} \delta_{cd}\nabla^{\lambda}\phi^{c} \nabla_{\lambda}\phi^{d}.
\end{eqnarray}
The function $\psi(\phi^{1}, \dots, \phi^{N})$ is restricted to nonminimal coupling dynamics, hence it vanishes in the minimal coupling case where $f = M^{2}_{\text{Pl}}$, and it shows also the differences between metric and purely affine gravity.

Given this overview on the multiple scalar fields coupled to gravity in its affine picture, we then turn to an interesting part about how to perform the transition from nonminimal to minimal couplings. 

\subsection{Transition to minimal coupling without geometric transformations}
\label{sec: transition}
In general, the transition from nonminimal to minimal coupling is essential since it brings the gravitational sector to a canonical form in a frame where the observed quantities are generally calculated. In metric theories this is achieved  by performing the so called conformal transformation where the metric tensor corresponding  to a Jordan frame is mapped to a new one referred  to as the  Einstein frame. However, in the absence of any metric, the purely affine actions  do not rely on this, but rather, only rescalings in field space associated with potential transformations would bring the gravitational sector to a canonical form. In fact,  action (\ref{action1}) could be brought to a more compact form  as
\begin{eqnarray}
\label{action-minimal}
S[\Gamma, \phi^{1}, \dots \phi^{N}]=
\int d^{4}x \frac{\sqrt{ | M^{2}_{\text{Pl}} R_{\mu\nu}(\Gamma)- \mathcal{G}_{ab}(\phi^{1}, \dots \phi^{N})\nabla_{\mu}\phi^{a} \nabla_{\nu}\phi^{b} |}}
{\tilde{V}(\phi^{1}, \dots, \phi^{N})},
\end{eqnarray}
where the original potential is rescaled as
\begin{eqnarray}
\label{potential-new}
\tilde{V}(\phi^{1}, \dots, \phi^{N})=
\left(\frac{M^{2}_{Pl}}{f(\phi^{1}, \dots, \phi^{N})}\right)^{2}V(\phi^{1}, \dots, \phi^{N}),
\end{eqnarray}
and the matrix, or the new metric of the $N$-dimensional field space $\mathcal{G}_{ab}$ is given by
\begin{eqnarray}
\label{field space metric-affine}
\mathcal{G}_{ab}(\phi^{1}, \dots \phi^{N})=
\frac{M^{2}_{Pl}}{f(\phi^{1}, \dots \phi^{N})} \delta_{ab}.
\end{eqnarray}
Now, the appearance of the factor $M^{2}_{\text{Pl}}$ translates the canonical form of the gravitational sector of the action. However the issue remains in the kinetic parts of the scalar fields. In fact, the field space metric is only conformal to flat which means that, in general, the components of the curvature tensor constructed from this new field space metric do not vanish identically. Generally, it is impossible that all $N$ scalars enjoy a canonical kinetic term, thus multiple scalar fields generically interact leading to entropy production.
\newline
On the other hand, it is very crucial to notice that the transition to minimal coupling dynamics has been achieved by transforming only the potential as (\ref{potential-new}) which is no longer sufficient when it comes to purely metric (like GR) or Palatini theories of gravity. In those standard theories, the gravitational part of the action is tightly glued to the metric tensor, thus the potential redefinition (\ref{potential-new}) must be performed in a new \enquote{geometric} frame (Einstein-frame) driven by a conformal transformation of the metric. Below we briefly outline some of various differences appearing in metric and Palatini theories compared to purely affine gravity presented in this paper:
\begin{enumerate}[(a)]
    \item In purely metric theories (GR and its modifications), the conformal transformation applied to the metric tensor is necessary for bringing a canonical form of the gravitational sector and brings extra terms to the quantity (\ref{field space metric-affine}) as \cite{kaiser-conformal}
\begin{eqnarray}
\label{field space metric-gr}
\mathcal{G}_{ab}(\phi^{1}, \dots \phi^{N})=
\frac{M^{2}_{Pl}}{2f(\phi^{1}, \dots \phi^{N})} \delta_{ab}
+
\frac{3M^{2}_{Pl}}{2f^{2}(\phi^{1}, \dots \phi^{N})} 
\frac{\partial f}{\partial \phi^{a}}\frac{\partial f}{\partial \phi^{b}}.
\end{eqnarray} 
The last term described by field derivatives arises due to the nonlinearity of Einstein-Hilbert action with respect to the metric tensor, which is generally unavoidable feature of purely metric theories. Thus, the presence of nonminimal interactions induce anisotropic stresses that are cleaned only via geometric (metric) conformal transformations.
   \item In Palatini formulation metric and affine connection, though independent, are both essential in forming the gravitational action \cite{demir-inflation}. The dynamical field here is the affine connection while metric gurantees the general covarince by forming scalar quantities by contraction. However, the gravitational equations must arise through variations with respect to both quantities. In this respect, since the metric is not dynamical the action is linear, thus the presence of nonminimal couplings generically do not generate anisotropic stresses like the case of metric theories a feauture that is shared with purely affine gravity. In this case it is known that the transition from nonminimal to minimal couplings generates a field-space metric not like (\ref{field space metric-gr}) but similar to (\ref{field space metric-affine}), however, the fact remains that a geometric (metric) conformal transformation is also required. We have to emphasize here that although it leads to field space metric similar to (\ref{field space metric-affine}), the Palatini gravity is not equivalent to purely affine gravity in various aspects. The main difference is that (\textit{i}) purely affine gravity is a metricless theory in which the purely affine actions must be completely independent of the metric and the latter arises only \textit{a posteriori} (\textit{ii}) when generated through the dynamical equations, the metric appears satisfying the compatibility condition (\ref{compatibility}) even in the case of nonminimal coupling at least in the present case where the connection is taken symmetric. Other crucial differences between Palatini and purely affine gravity would certainly arise when the connection and the Ricci tensor are left freely asymmetric.         
   \item Concerning the present approach, as shown from the detailed study carried out in the last section, the Purely affine gravity stands on connection solely from which emerges a metric tensor that is not imposed from scratch. The advantage here is providing a unique geometric frame (unique metric) for both nonminimally and minimally coupled scalar fields. In other words matter fields described by scalar fields live and propagate in a unique geometric frame, and in this respect FRW background and associated quantities such as Hubble parameter are not affected by switching between different couplings.
\end{enumerate}

\subsection{Background field dynamics}
\label{sec: background field dynamics}
Before tackling the evolution of the perturbations, let us first examine the dynamics of the homogeneous parts of the scalar fields in a spatially flat Friedmann-Robertson-Walker (FRW) universe. It is from the homogeneous background fields that one imposes the slow roll conditions which finally provide us with solutions to the flatness and horizon problems.

In what follows, for simplicity we will take the Planck mass $M^{2}_{\text{Pl}}=1$ which can be easily recovered in practice.

Taking the scalar fields as homogeneous, $\phi \sim \varphi(t)$, the generalized energy momentum tensor (\ref{energy-momentum tensor}) would simply split into a generalized energy density and pressure given as
\begin{eqnarray}
\label{density and pressure}
&&\rho = \frac{1}{f} \left( \frac{1}{2} \dot{\varphi}^{a} \dot{\varphi}^{a} + \frac{V}{f} \right) \quad \text{and} \quad
 \mathcal{P} = \frac{1}{f} \left( \frac{1}{2} \dot{\varphi}^{a} \dot{\varphi}^{a} - \frac{V}{f} \right), 
\end{eqnarray}
where the nonminimal coupling function and the potential are evaluated at the background, $f=f(\varphi)$ and $V=V(\varphi)$. We will keep this notation when treading the field fluctuations later.

The gravitational field equations (\ref{eom-gravity}) are easily adapted for the flat FRW spacetime leading to  
\begin{eqnarray}
\label{hubble parameter}
3H^{2}=
\frac{1}{f}
\left( \frac{1}{2} \dot{\varphi}^{a} \dot{\varphi}^{a}
+\frac{V}{f}
 \right)
\end{eqnarray}
and 
\begin{eqnarray}
\dot{H} + H^{2} =
-\frac{1}{3f}
\left( \frac{1}{2} \dot{\varphi}^{a} \dot{\varphi}^{a}
- \frac{V}{f}
 \right)
\end{eqnarray}
The same for the evolution equation (\ref{eom-field dynamics}) which takes the form
\begin{eqnarray}
\label{eom in frw}
\ddot{\varphi}^{a} +
3H\dot{\varphi}^{a} +\frac{1}{f} V_{,a} - 3 (\dot{H} + 2H^{2}) f_{,a}
-\frac{1}{f} \dot{\varphi}^{b} \dot{\varphi}^{b} f_{,a}=0.
\end{eqnarray}
Here $H=\dot{a}/a$ is the Hubble parameter in terms of the scale factor $a(t)$. For the ease of notation we have used the sign \enquote{comma} to refer to derivatives with respect to the scalar fields. We have also omitted the $\delta_{ab}$ symbol leaving only repeated indices for summation convention. Equation (\ref{eom in frw}) can also be derived from the conservation of the total energy-momentum tensor (\ref{energy-momentum tensor}) which takes the common form 
\begin{eqnarray}
\dot{\rho}+3H(\rho+\mathcal{P})=0  
\end{eqnarray}
It is easy to notice the differences from the metric gravity in the case of both minimal and nonminimal couplings. First of all, the minimal coupling limit ($f=1$) is equivalent to that in metric gravity where the set of equations (\ref{density and pressure})-(\ref{eom in frw}) are reduced to the standard cosmology equations in the presence of single field. In the nonminimal case where the function $f$ is a field-dependent, the energy density and pressure as well as the potential are modified by a multiplicative factor $f^{-1}$ compared to the minimal case. However, when compared to the nonminimal case of metric gravity we realize a crucial difference, in addition to the factor $f^{-1}$ we notice here the absence of the first and second time-derivative of the function $f$ in the energy density and pressure due to the absence of anisotropic terms in the energy-momentum tensor (\ref{energy-momentum tensor}). In other words, the differences between minimal and nominimal couplings dynamics in affine gravity arises only through simple factors not in additional terms.       

\section{Field fluctuations and entropy production}
 \label{sec: field fluctuations and entropy perturbation}
\subsection{Scalar perturbations and anisotropic stress-free dynamics}
In every model of inflation, inhomogeneities in the scalar fields are of great importance since they lead to curvature perturbations which in turn provides the measure of gauge invariant primordial perturbations acting as seeds for structure formation. In the following, we will follow the standard way of deriving the scalar perturbations dynamics from the equations of motion which are in this case summarized in (\ref{eom-gravity})-(\ref{psi}).  First we expand the fields around a homogeneous backgrounds $\varphi^{a}$ as
\begin{eqnarray}
\phi^{a} = \varphi^{a}(t)
+\delta \phi^{a}(t, \vec{x}), 
\end{eqnarray} 
where the first term satisfies the equations of motion of the last section, and the last term represents the multiple fields fluctuations.

We then impose deviations from FRW spacetime that would represent a perturbed metric in which $g_{00}=-(1+2\Phi)$ and $g_{ij}=a^{2}\delta_{ij}(1-2\Psi)$, where $\Psi$ and $\Phi$ are space and time dependent scalar potentials. Thus, up to first order, the generalized energy momentum tensor perturbations lead to a generalized energy density and pressure fluctuations as\footnote{Remember that we are taking $M^{2}_{\text{Pl}}=1$ for brevity.} 
\begin{eqnarray}
\label{density fluctuations}
&&\delta \rho =
\frac{1}{f} \left(\dot{\varphi}^{a}\delta \dot{\phi}^{a} - \dot{\varphi}^{a}\dot{\varphi}^{a} \Phi
+ \frac{1}{f} V_{,a} \delta\phi^{a}
 \right)
 -\frac{1}{f^{2}} \left( \frac{1}{2}  \dot{\varphi}^{b}\dot {\varphi}^{b} f_{,a} +\frac{2V}{f} f_{,a} \right) \delta \phi^{a}, \\
&&\delta \mathcal{P} =
\frac{1}{f} \left(\dot{\varphi}^{a}\delta \dot{\phi}^{a} - \dot{\varphi}^{a}\dot{\varphi}^{a} \Phi
- \frac{1}{f} V_{,a} \delta\phi^{a}
 \right)
 -\frac{1}{f^{2}} \left( \frac{1}{2}  \dot{\varphi}^{b}\dot {\varphi}^{b} f_{,a} - \frac{2V}{f} f_{,a} \right) \delta \phi^{a}
\label{pressure fluctuations}
\end{eqnarray}
Thus, the time-time part of the gravitational equations (\ref{eom-gravity}) reads
\begin{eqnarray}
\label{00 component}
3H(\dot{\Psi} + H\Phi)+\frac{1}{a^{2}}\vec{\nabla}^{2}\Psi =
-\frac{1}{2f} \left(\dot{\varphi}^{a}\delta \dot{\phi}^{a} - \dot{\varphi}^{a}\dot{\varphi}^{a} \Phi
+ \frac{1}{f} V_{,a} \delta\phi^{a}
 \right) 
+\frac{1}{2f^{2}} \left( \frac{1}{2} \dot{\varphi}^{b}\dot {\varphi}^{b} f_{,a} +\frac{2V}{f} f_{,a} \right) \delta \phi^{a}
\end{eqnarray}
whereas the time-space part leads to 
\begin{eqnarray}
\label{0 i component}
\dot{\Psi} + H\Phi =
\frac{1}{2f} \dot{\varphi}^{a} \delta \phi^{a}.
\end{eqnarray}
Note that these derivatives appear in the expansion of the potential and the nonminimal coupling function $f$ around the background fields, i.e, $f(\varphi^{a}+ \delta \phi^{a}) \simeq f(\varphi^{a})+f_{,b}(\varphi^{a}) \delta \phi^{b}+ \mathcal{O}(\delta \phi^{a}\delta \phi^{b})$, and the same for the potential.

The second important evolution equation which generically describes a constraint on the scalar potentials $\Psi$ and $\Phi$ could be derived easily from the spacial part of the gravitational equations (\ref{eom-gravity}) and reads 
\begin{eqnarray}
\label{psi-phi}
\nabla_{i} \nabla_{j} \left( \Psi  - \Phi \right)=0 \quad (\text{for} \quad i \neq j).
\end{eqnarray}
This interestingly shows that even multiple fields, coupled nonminimally to gravity, do not contribute to \textit{anisotropic stress}. In other words, the evolution equation (\ref{psi-phi}) which generically holds in the minimal coupling case is \textit{conserved} when every possible nonminimal interaction is present. In metric gravity however, this is no longer the case. In fact, the nonlinearity of the actions would result in the presence of a generalized energy momentum tensor from which arises an anisotropic stress, and finally the right hand side of (\ref{psi-phi}) would not vanish \cite{kaiser0, kaiser1}. The appearance of the anisotropic term in the nonminimal coupling dynamics of GR means that the Bardeen potentials totally differ.

In the standard cosmological model, baryons and cold dark matter do not contribute anisotropic stress since they are successfully approximated to perfect fluids. Photons and neutrinos on the other hand could in principle contribute anisotropic stress when they have considerable quadrupole moments. While photons contribute less, collisionless neutrinos however have appreciable quadrupole moments during the radiation dominated era \cite{dodelson}. In the case of scalar fields and mainly minimally coupled ones, one can considerably simplify the equations of motion and show that to first order in perturbations the spacial part of the energy momentum tensor is proportional to $\delta^{i}_{j}$, thus do not contribute any source to the right hand side of (\ref{psi-phi}). This happens in both GR and pure affine gravity\footnote{In the case of minimal coupling, affine gravity, though generally different, leads to the same dynamics as GR, however, deviations from GR become crucial when the fields are nonminimally coupled \cite{affine-inflation}.} when the fields are minimally coupled to gravity. In our case, the reason for which the evolution relation (\ref{psi-phi}) is not altered by the nonminimal coupling is that one could move simply from different couplings by simply redefining the scalar fields and not the metric, an important feature in affine gravity which has been described in section (\ref{sec: transition}). In GR, however, the field redefinition is necessarily followed by a metric conformal mapping which in principle alters relation (\ref{psi-phi}).      

Condition (\ref{psi-phi}) which is equivalent to $k^{2}( \Psi  - \Phi)=0$ in Fourier space means simply that we end up with only one scalar potential $\Psi  = \Phi$ which finally simplifies the the above evolution equations.

\subsection{Sources of entropy perturbations}    
An important scalar quantity in cosmological perturbations is the gauge-invariant \textit{curvature perturbation} on a uniform-density hypersurfaces \cite{dodelson}
\begin{eqnarray}
\label{zeta}
\zeta \equiv - \Psi -\frac{H}{\dot{\rho}} \delta \rho,
\end{eqnarray}
where $\rho$ is the energy density which is caused by scalar fields in the present case. 

An interesting feature of this quantity is that it remains constant outside the horizon for \textit{adiabatic} matter perturbations. In the case of single scalar field minimally coupled to gravity, it can  easily be shown that the perturbation (\ref{zeta}) does not evolve outside the horizon, i.e, when $k \ll aH$. The reason is simply that minimally coupled slowly rolling single field does not contribute \enquote{non-adiabatic} pressure on super-horizon scales, i.e
\begin{eqnarray}
\delta \mathcal{P}_{nad} \equiv \delta \mathcal{P} -\frac{\dot{\mathcal{P}}}{\dot{\rho}} \delta \rho =0 \quad \text{for} \quad k \ll aH
\end{eqnarray} 
The quantity $\delta \mathcal{P}_{nad}$ refers to the non-adiabatic pressure, and any source that contribute a nonzero value to this quantity would imply an evolving curvature perturbation on cosmologically interesting length scales. This in turn ends up with producing considerable \textit{entropy perturbations}. In the following, our goal is to examine these entropy perturbations sourced by multiple scalar fields nonminimally coupled to (affine) gravity. Hence, every source of entropy perturbation shall appear through every possible nonzero term that forms the non-adiabatic pressure calculated from the \enquote{generalised} energy density and pressure (\ref{density and pressure}). 

Energy density and pressure with their associated fluctuations (\ref{density fluctuations}) and (\ref{pressure fluctuations}) lead to
\begin{eqnarray}
\delta \mathcal{P}_{nad} =
-\frac{2\dot{\varphi}^{a}V_{,a}}{3H \dot{\varphi}^{c} \dot{\varphi}^{c} f}
\delta \rho
+ \frac{4 V \dot{\varphi}^{a}f_{,a}}{3H \dot{\varphi}^{c} \dot{\varphi}^{c} f^{2}} \delta \rho
- \frac{2 V_{,a}\delta \phi^{a} }{f^{2}}
+\frac{4 V f_{,a} \delta \phi^{a}}{f^{3}},
\end{eqnarray}
where we have used $\dot{\rho}= -3H(\rho +\mathcal{P})$.

We then add and subtract the term $3H\dot{\varphi}^{b}\delta \phi^{b}/f$ to obtain
\begin{eqnarray}
 \delta \mathcal{P}_{nad} =&&
 -\frac{2\dot{\varphi}^{a}V_{,a}}{3H \dot{\varphi}^{c} \dot{\varphi}^{c} f}  
 \left( \delta \rho +\frac{3H}{f} \dot{\varphi}^{b}\delta \phi^{b} \right)
 + \frac{4 V \dot{\varphi}^{a}f_{,a}}{3H \dot{\varphi}^{c} \dot{\varphi}^{c} f^{2}} 
 \left( \delta \rho +\frac{3H}{f} \dot{\varphi}^{b}\delta \phi^{b} \right) \nonumber \\
 &&
 +\frac{2 \dot{\varphi}^{a}V_{,a} \dot{\varphi}^{b} \delta \phi^{b}}{f^{2} \dot{\varphi}^{c} \dot{\varphi}^{c}}
 - \frac{2 V_{,a}\delta \phi^{a} }{f^{2}}
 - \frac{4 V \dot{\varphi}^{a}f_{,a} \dot{\varphi}^{b} \delta \phi^{b}}{f^{3} \dot{\varphi}^{c} \dot{\varphi}^{c}}
 +\frac{4 V f_{,a} \delta \phi^{a}}{f^{3}}.
\end{eqnarray}
The term in parenthesis is the generalized gauge-invariant comoving density perturbation and can be obtained easily by combining the evolution equations (\ref{00 component}) and (\ref{0 i component}) which yield
\begin{eqnarray}
\delta \rho +\frac{3H}{f} \dot{\varphi}^{b}\delta \phi^{b} = -2\frac{k^{2}}{a^{2}}\Psi,
\end{eqnarray}
where $k$ is the wave vector (momentum) that comes out of $\vec{\nabla}^{2}\Psi$ in Fourier space. 

Finally, the generalized non-adiabatic pressure reads 
\begin{eqnarray}
\label{p nad final}
 \delta \mathcal{P}_{nad} =&&
  \underbrace{ \frac{4 H \dot{\varphi}^{a}V_{,a}}{3 \dot{\varphi}^{c} \dot{\varphi}^{c} f} \left(\frac{k}{aH} \right)^{2} \Psi}_{\text{Suppressed term}} 
-\underbrace{ \frac{8 H V \dot{\varphi}^{a}f_{,a}}{3 \dot{\varphi}^{c} \dot{\varphi}^{c} f^{2}} \left(\frac{k}{aH} \right)^{2} \Psi }_{\text{Suppressed term}}  \nonumber \\
&&
-\frac{2V_{,a}}{f^{2}}\left[ \delta \phi^{a} -\frac{\dot{\varphi}^{a}}{\dot{\varphi}^{c} \dot{\varphi}^{c}} \dot{\varphi}^{b} \delta \phi^{b} \right]
+\frac{4V f_{,a}}{f^{3}}\left[ \delta \phi^{a} -\frac{\dot{\varphi}^{a}}{\dot{\varphi}^{c} \dot{\varphi}^{c}} \dot{\varphi}^{b} \delta \phi^{b} \right]
\end{eqnarray}
While the first two terms are suppressed on supper-horizon scales the above non-adiabatic pressure remains nonzero due to the presence of two terms
\begin{eqnarray}
\delta \mathcal{P}_{nad} \supset
 \delta \mathcal{P}_{nad}^{\text{multiple}}  + \delta \mathcal{P}_{nad}^{\text{non-min}},
\end{eqnarray}
which represent the two possible and distinct sources of entropy perturbations and they are as follows:
\begin{enumerate}
\item \textit{Source from multiple fields}

The first source is induced by multiple fields and is described by the first (not suppressed) term in (\ref{p nad final}) 
\begin{eqnarray}
\label{first source}
\delta \mathcal{P}_{nad}^{\text{multiple}} =
-\frac{2V_{,a}}{f^{2}}\left[ \delta \phi^{a} -\frac{\dot{\varphi}^{a}}{\dot{\varphi}^{c} \dot{\varphi}^{c}} \dot{\varphi}^{b} \delta \phi^{b} \right].
\end{eqnarray}
In fact, one can easily verify that this source vanishes for single (nonminimally or minimally) coupled scalar field, i.e, when $N=1$ ($\phi^{a} \equiv \phi$) for both cases $f = constant$ (minimal) or $f \neq constant$ (nonminimal)
\begin{eqnarray}
\delta \mathcal{P}_{nad}^{\text{multiple}}
\xrightarrow{\text{For single scalar field}} 0.
\end{eqnarray}
However, for instance, two scalar fields $\phi^{a} =(\phi, \chi)$ would produce
\begin{eqnarray}
\delta \mathcal{P}_{nad}^{\text{multiple}} =
-\frac{2 \dot{\phi} \dot{\chi}}{(\dot{\phi}^{2} + \dot{\chi}^{2})f(\phi, \chi)} \left( \dot{\chi} V_{,\phi} - \dot{\phi} V_{,\chi} \right)
\left(\frac{\delta \phi}{\dot{\phi}} -\frac{\delta \chi}{\dot{\chi}} \right),
\end{eqnarray} 
for a general coupling function $f(\phi, \chi)$ including a constant (minimal coupling). 

This does not vanish indeed. Thus, entropy perturbation is a generic feature of multifields, and as we have shown here, it occurs also in pure affine gravity. In general, this feature has been interpreted by the fact that the presence of multiple fields would lead to multiple trajectories in the phase space where the vacuum fluctuations that are stretched to super-Hubble scales would then inevitably include non-adiabatic perturbation \cite{wands0, wands1, wands2}. It is clear that this contribution must not be suppressed in the long wavelength limit.
   
\item \textit{Source from nonminimal coupling}

The second source of entropy perturbation in this framework is related to the nonminimal coupling and it is described by the last term in (\ref{p nad final}) or
\begin{eqnarray}
\label{second source}
\delta \mathcal{P}_{nad}^{\text{non-min}}=
\frac{4Vf_{,a}}{f^{3}}\left[ \delta \phi^{a} -\frac{\dot{\varphi}^{a}}{\dot{\varphi}^{c} \dot{\varphi}^{c}} \dot{\varphi}^{b} \delta \phi^{b} \right].
\end{eqnarray}
This quantity clearly vanishes for a constant $f$, which is the case of minimal coupling dynamics. Thus, the first remark is that when the coupling is minimal, only contributions from multifields (the previous source) induce entropy perturbations. 

Another interesting feature of this source is that like $\delta \mathcal{P}_{nad}^{\text{multiple}}$, it also vanishes for single scalar. This interestingly means that nonminimal couplings have effects on the entropy perturbations only for the case of more than one scalar field. This does not hold in metric gravity where even a single field can source an entropy perturbation if it is nonminimally coupled \cite{kaiser0, kaiser1}. 

Unlike the present case, in metric gravity non-adiabatic pressure induced by nonminimal coupling appears as a contribution of several separate terms due to the complicated form of the energy-momentum tensor (see the discussion below equation (\ref{energy-momentum tensor})). In particular, terms that generate anisotropic pressure have their effects even when only one single field is considered, the fact that prevents the non-adiabatic pressure from vanishing for a single field as well. We note here   that some confusions concerning the notion of adiabaticity may arise. In fact, it is known that nonminimal coupling dynamics can be easily transformed to minimal coupling dynamics (to Einstein frame) without altering the physics. In Einstein frame however, single scalar cannot contribute to any source of non-adiabatic pressure that are not suppressed on super-horizon scales, hence, it does not generate any entropy perturbation. In other words, while entropy perturbation is suppressed in one frame (Einstein frame), it is generated in the other one (Jordan frame). This confusion results from the conformal transformation of the metric which is necessary for switching from one to another frame.        
 
In our framework, based on affine gravity, those conformal frames arising from conformal transformation are not present. As we have seen in Sec. \ref{sec: transition}, the transition to minimal coupling is made by performing only scalar fields redefinition. The metric tensor in this sense is unique for both couplings (minimal and nonminimal), it has been generated dynamically from a pure affine action that does not refer to any metric to transform. This metric remains the same when switching to minimal coupling dynamics. Thus, in this picture the notion of adiabaticity is invariant under field redefinition. While multiple fields induce entropy perturbations in both nonminimal and minimal coupling cases, single scalar field does not in both cases as well.
\end{enumerate}
The gauge-invariant curvature perturbation (\ref{zeta}), a crucial quantity in primordial cosmology, represents a measure of the primordial perturbations which lead to fluctuations of the temperature in the cosmic microwave background and finally manifest as seeds for structure formation. Its conserved character, which is a generic but crucial feature in the most known models of inflation (particularly with single fields), turns out to be altered in the presence of multifields.  In fact, one can show that the evolution (time dependence) of the curvature perturbation is generically proportional to the non-adiabatic pressure \cite{wands0, wands1, wands2}  
\begin{eqnarray}
\dot{\zeta}  = - \frac{H}{\rho + \mathcal{P}} \delta \mathcal{P}_{nad} + \text{terms suppressed by} \, \left(\frac{k}{aH} \right)^{2}.
\end{eqnarray}
This relation can be derived for every conserved energy-momentum tensor including our present case in which it leads to
 \begin{eqnarray}
 \label{zeta-dot}
\dot{\zeta} \supset - \frac{Hf}{\dot{\varphi}^{a}\dot{\varphi}^{a}} \left( \delta \mathcal{P}_{nad}^{\text{multiple}}  + \delta \mathcal{P}_{nad}^{\text{non-min}} \right).
\end{eqnarray}
From the conclusions drawn above, we may safely say that unlike in metric gravity, here it is sufficient to consider only one single scalar field and one then recovers the conservation of the curvature perturbation ($\dot{\zeta} = 0$). In the single-field case then, nonminimal coupling to affine gravity does not break the conservation law of the curvature perturbation. At the theoretical level, this is a new and important feature compared to inflation in the context of metric gravity, and it would be interesting to be able to  probe it at the observational level. Unfortunately, for the time being there is no observational strategy in which only the conservation of the curvature perturbation would provide us with a clue in distinguishing between different theories of gravity.

In the case of multiple fields, however, $\dot{\zeta} \neq 0$ and deviations from zero result from the contributions of the above sources of non-adiabatic pressure induced by the presence of more than one scalar. Entropy perturbation modes or as commonly called \enquote{isocurvature} perturbations can contribute to both power spectrum and bispectrum if they survive until the recombination era. Possible \enquote{cross-correlation} between adiabatic and isocurvature would also lead to production of mixed bispectra \cite{komatsu, langlois1, langlois2}. Hence on cosmological scales isocurvature perturbations, if any, may be constraints by observations, for instance by analyzing the angular power spectrum of the cosmic microwave background \cite{planck2018}.  

We conclude by mentioning that the perturbation that is constrained by observations is the density perturbation at the era of primordial nucleosynthesis. Thus, the presence of isocurvature (non-adiabatic) perturbations\footnote{At first sight it seems that from (\ref{zeta-dot}), one can make $\dot{\zeta} =0$ (adiabatic perturbations) if $f \propto V^{1/2}$. Unfortunately, this constrain affects at the first place the dynamics of the inflaton since it also clears away the effects of the potential in (\ref{hubble parameter}) leaving only kinetic terms of the inflaton that do not allow slow roll conditions.} during inflation does not necessarily imply its presence at later times \cite{weinberg}. Indeed, whether the non-adiabatic modes remain non-adiabatic or not, may depend on the process of reheating (if any) which is still yet to be understood \cite{wands0}.

\section{Summary and conclusion}
The last few decades were remarkable for cosmology where even the physics of the early universe becomes accessible to high-precision observations. Lot of efforts have been devoted in parallel to different theoretical models for the early universe particularly those of inflation with the aim of coming up with one successful and convincing model that fits the accurate data. Among the inflationary models that gained much attentions recently are those of nonminimally interacting multiple fields \cite{kaiser0, kaiser1, sasaki} (see also \cite{attractor} as example for attractor behavior in multifields-inflation). Indeed, while realistic models of elementary particles typically include many scalars, quantum field theory in curved spacetime generically requires nonminimal couplings for the scalar fields.     

 However when more than one scalar are present crucial changes arise (compared to the case of single field) not only in the dynamics but also in the perturbation itself that is generated during inflation. In fact, the fields may interact and cause significant non-adiabatic (entropy) perturbations that typically are not suppressed on super-horizon scales. Studies of these isocurvature perturbations and their possible detection through the Cosmic Microwave Background anisotropies and polarizations is at the heart of every serious work on multiple-fields inflation \cite{wands0, wands1, wands2, komatsu, langlois1, langlois2, tent}. 

In this paper we have presented a general framework in which multiple fields are considered to drive inflation but in the context of purely affine gravity rather than in general relativity. Indeed, we believe that not only the type of the fields are important in the very early universe but also the approach to gravity can play a crucial role. We have started with a metric-less action from which the metric tensor itself arises through the equation of motion. The primary goal was to investigate the possible sources of non-adiabatic pressure that cause entropy perturbations not only from the presence of multiple fields (which is a generic feature) but also due to the nonminimal interactions.

The scalar perturbations have shown that there must be two distinct sources of non-adiabaticity, one is the familiar source arising from multiple fields and the other one is related to nonminimal coupling. Although the two types of sources are expected as in metric gravity, here the source that arise from nonminimal couplings vanishes when only single scalar is considered. In other words, entropy perturbations are there simply because there are multiple fields. The later remark leads us to raise the frame-issue encountered in metric gravity \cite{sasaki}. In fact, detailed calculations made in Jordan frame showed that even when only a single scalar is considered there will be still an entropy perturbations that survive when the field is nonminimally coupled \cite{kaiser0, kaiser1}. We know however that in Einstein frame where the inflaton is minimally coupled to gravity, non-adiabatic perturbations are suppressed in the long wavelength limit leading to adiabatic curvature perturbations. In other words, although the conformal frames are physically equivalent, we notice that while the curvature perturbation is adiabatic in one frame, it is non-adiabatic in the other one, thus the change of frames confuses the notion of adiabaticity in metric theories of gravity. In the present framework it is clear that this frame issue is not encountered in affine gravity where the origin of the confusion which is the confromal transformation is not present. 

It is only possible future more precise measurements, for instance of the power spectrum of Cosmic Microwave Background temperature and anisotropies that will show whether there were really isocurvature (entropy) perturbations that are generated during inflation and survived until recombination. This must be investigated along with a specific model based on the present framework \cite{azri-nasri-in-progress}.      

\label{sec: conclusion}

\begin{acknowledgments}
H. A is grateful to Durmu\c{s} Demir for useful  discussion.
\end{acknowledgments}

\bibliographystyle{apsrev4-1.bst}
\bibliography{references}

\end{document}